# ESTIMATING FIRE WEATHER INDICES VIA SEMANTIC REASONING OVER WIRELESS SENSOR NETWORK DATA STREAMS

Lianli Gao[1], Michael Bruenig[2], Jane Hunter[3]

[1] School of ITEE, The University of Queensland, Brisbane, QLD 4072, Australia
[2] CSIRO ICT Centre, Brisbane, QLD4069, Australia
[1] School of ITEE, The University of Queensland, Brisbane, QLD 4072, Australia

## ABSTRACT

*Wildfires are frequent, devastating events in Australia that regularly cause significant loss of life and widespread property damage. Fire weather indices are a widely-adopted method for measuring fire danger and they play a significant role in issuing bushfire warnings and in anticipating demand for bushfire management resources. Existing systems that calculate fire weather indices are limited due to low spatial and temporal resolution. Localized wireless sensor networks, on the other hand, gather continuous sensor data measuring variables such as air temperature, relative humidity, rainfall and wind speed at high resolutions. However, using wireless sensor networks to estimate fire weather indices is a challenge due to data quality issues, lack of standard data formats and lack of agreement on thresholds and methods for calculating fire weather indices. Within the scope of this paper, we propose a standardized approach to calculating Fire Weather Indices (a.k.a. fire danger ratings) and overcome a number of the challenges by applying Semantic Web Technologies to the processing of data streams from a wireless sensor network deployed in the Springbrook region of South East Queensland. This paper describes the underlying ontologies, the semantic reasoning and the Semantic Fire Weather Index (SFWI) system that we have developed to enable domain experts to specify and adapt rules for calculating Fire Weather Indices. We also describe the Web-based mapping interface that we have developed, that enables users to improve their understanding of how fire weather indices vary over time within a particular region. Finally, we discuss our evaluation results that indicate that the proposed system outperforms state-of-the-art techniques in terms of accuracy, precision and query performance.*

## 1. INTRODUCTION

Wildfires have been responsible for some of the most devastating natural disasters in Australia and are estimated to cause damage with an average annual cost of $77million [1]. Fire weather indices play a significant role in issuing warnings and in estimating the level of difficulty associated with a potential wild fire/bushfire [2]. The most widely used and accepted systems are the McArthur Forest Fire Danger Index (used in Australia) and the Canadian Fire Weather Index

(used in North America and the Australia Bureau of Meteorology (BoM)) [2]. Both indices are calculated by making use of three weather parameters: wind speed, relative humidity and temperature. Although these two fire weather index systems are the most robust and widely adopted, they have some limitations. For example, most existing implementations use weather parameters collected from widely distributed sensor nodes, tens of kilometers apart. Hence, the collected data is not dense enough to estimate fire weather index accurately for a specific region. Moreover, the fire danger maps are typically only updated once per day. Hourly variations in bushfire risk over the course of 24 hours are not possible. For example, the Canadian Fire Weather Index takes the noon Local Standard Time (LST) values of weather parameters as input. The potential cost of imprecise information can be very significant in decision making associated with hazard evacuation plans and fire-fighting operations. Therefore, there is an urgent need for developing more accurate methods for estimating fire weather indices, with higher spatio-temporal resolutions.

In recent years, the number of wireless sensor networks (WSNs) deployed in different environments has rapidly expanded due to the decreasing cost, size and reliability of micro-sensor technologies. Sensors are being used to monitor safety and security of buildings and spaces [3,4], to measure humans' physical, physiological, psychological, cognitive and behavioral processes [5] and to capture observations and measurements of the environment parameters [6]. In recent years, WSNs consisting of coordinated autonomous sensors have been deployed to monitor forest physical parameters (air temperature, relative humidity, wind speed, leaf wetness, air pressure, wind direction, solar radiation and so on) [6,7,8]. As a result, an avalanche of raw sensor network data streams about forest environments has been collected which provides a valuable research platform for scientists or researchers to study or understand the micro-climate and associated fire weather indices within a focused area [6].

However, estimating fire weather indices from WSNs streams is a very challenging problem. Firstly, WSN data streams are often incomplete or imprecise due to the fading signal strength, hazard node faults, inaccuracies of measurement, and limited energy and wireless bandwidth [9,10]. Moreover, the large volumes of complex, numerical and unstructured sensor data streams being generated, is a major challenge to process in real-time or near-real-time. In addition, heterogeneous, non-standard infrastructure, and poor data representation have resulted in many sensor data streams being locked inside specific proprietary applications and inaccessible to the wider community. To date, only a few studies such as Semantic Sensor Web [11] have focused on addressing the limitations of raw sensor data streams by annotating sensor data with semantic metadata to improve their spatial, temporal and semantic meaning and their potential interoperability and re-usability. The Semantic Sensor Web research also demonstrated the application of semantic reasoning rules to derive new knowledge from semantically annotated sensor data. However, information processing and analysis of sensor network streams remain in its infancy with much related effort focusing on the provision of platforms to support more efficient sensor-based applications or the improvement of sensor-based data management, sensor network configuration and sensor communication protocols [12,13].

The aim of the research described in this paper is to develop a Semantic Fire Weather Index (SFWI) system which combines data pre-processing techniques (e.g., outlier detection and semantic annotation), with semantic reasoning technology and domain expert knowledge to estimate fire weather indices from WSNs data streams collected from a network deployed in the Springbrook region of South east Queensland [6]. More specifically, the system is designed to satisfy the following objectives and user requirements:

- To detect and remove the outliers within the raw sensor data streams to improve the quality of the data streams. Following removal of outliers, the cleaned sensor data stream are converted to RDF triples - which improves the quality, sharing-ability and reusability of the sensor data streams.
- To develop a Fire Weather Index ontology (in OWL, Web Ontology Language) to represent different levels of fire weather danger ratings. These fire weather danger ratings based on input from the meteorological domain experts.
- To define First-Order-Logical inference rules for estimating fire weather indices, and then convert these rules into SPARQL inference rules by using SPARQL [14] and OWL ontologies.
- To design an efficient storage technique for storing, querying and retrieving large volume of sensor observation RDF triples efficiently. A multiple repository storage method is evaluated.
- To develop an inference algorithm to infer the fire weather indices (FWIs) for a specific region and a given time period by combining SPARQL inference rules with an Inverse Distance Weighting [15] based approach. This combined approach enables accurate spatial distributions of FWIs to be inferred from point data.
- To develop a set of Web services that enable users to search, explore and visualize fire weather indices within a time period for a specific region - using Google Earth, timeline and pie chart visualizations.

The remainder of the paper is structured as follows. In section 2, we briefly discuss related work. Section 3 presents our methodology. Section 4 describes the data pre-processing steps, which includes detecting and removing outliers, annotating data streams with terms from a set of OWL ontologies, and storing the RDF triples in optimized RDF storage. In section 5, we describe how we combine meteorologists' knowledge with semantic reasoning technology to infer accurate fire weather indices. Section 6 provides details about the system's technical architecture, functionality and the user interface. Section 7 provides detailed information about the evaluation process and results. Lastly, we discuss the future work and draw conclusions.

## 2. RELATED WORK

Using satellite telemetered data to detect and forecast forest fires is the traditional approach and still the predominant method. For example, a wildfire monitoring service was proposed to show how satellite images, ontologies and linked geospatial data can be combined for wildfire monitoring [16,17]. This approach processed satellite images to detect pixels where the fire may exist and then representing satellite image metadata, knowledge extracted from satellite images and auxiliary geospatial datasets encoded as linked data.

However, it has been proven that using WSNs to detect and forecast forest fires provides more timely and higher density data than using traditional satellite telemetered data [18]. Hence a number of recent efforts have focused on monitoring forest fires using WSNs [13,18,19,20,21]. Generally, these past research efforts can be classified into two categories: WSN level and WSN data analysis level. At the WSN level, previous research has focused on improving the WSN configurations, deployment, communication protocols, and hardware devices to enable more efficient fire detection and monitoring. For example, Aslan *et al.* [13] have proposed a framework consisting of four main components: an approach for deploying sensor nodes, an architecture for the sensor network for fire detection, an intra-cluster communication protocol, and an inter-cluster communication protocol. The aim is to improve the energy efficiency, support early detection and accurate localization, enable forecast capability, and adapt sensor networks to harsh environments. Fernández-Berni *et al.* [21] have been working on early detection of forest fires using a vision-enabled WSN. This work includes a vision algorithm for the detection of smoke and a low-power smart imager to stream images. They integrated these two components to generate a prototype vision-enabled sensor network node.

At the WSN data analysis level, researchers are mainly working on investigating the best sensor combinations (light, air temperature, air pressure, wind speed, wind direction, soil moisture, leaf wetness, relative humidity, rainfall, smoke etc.) and on developing more advanced algorithms (clustering, summaries, threshold, statistical modeling, neural network, threshold values, Dempster-Shafer theory based algorithm etc.) [13,20,22,23] to improve fire hazard detection and monitoring. For example, Diaz-Ramirez *et al.* [20] proposed two algorithms for detecting forest fires from WSNs. The first algorithm is a threshold-based method which takes temperature, humidity and light as input, while the second algorithm is a Dempster-Shafer theory based algorithm which only takes temperature and humidity as input.

The FireWatch [24] system was proposed to overcome the limitations of the traditional satellite and camera-based systems by integrating WSN technologies, computer-supported cooperation work (CSCW) and a Geographic Information System (GIS). This system was designed to detect forest fires using WSNs but not to support fire hazard predictions.

At present, only a few approaches have directly focused on calculating fire weather index from the WSN data streams [25,26]. For example, Sabit *et al.* [25] have presented approaches to generate micro-scale estimates of the Fire Weather Index from WSN data streams – but they do not use Semantic Web technologies.

Other noteworthy recent effort [27,28,29] proved that the integration of Semantic Web technologies and sensor networks can offer more to users and enable the development of environmental decision support systems, e.g., flood response planing system. Moreover, Sheth *et al.* [27] have proposed the Semantic Sensor Web to address integration and communication problems between networks by annotating sensor data with semantic metadata to improve their spatial, temporal and semantic meaning. It also demonstrates how rules can be applied to derive additional knowledge from semantically annotated sensor data. In our work, we adopt a similar semantic annotation or mark-up approach to enrich WSN data streams with semantic metadata to improve their data quality, share-ability and reusability. The primary difference between our work and Sheth's is that we apply and evaluate this approach in the context of estimating fire weather indices at a fine spatial and temporal scale.

## 3. METHODOLOGY
### 3.1 Case Study

The Gold Coast Springbrook National Park is one of Queensland's five World Heritage listed areas and covers 6,197 hectares restored from agricultural grassland to native rainforest vegetation. Located about 96 km south of Brisbane in the state of Queensland, it is a place of exceptional natural beauty and ecological importance. The Springbrook WSN project [6] being undertaken by CSIRO, Queensland Department of Environment and Resource Management, and Australia Rainforest Conservation Society has employed more than 180 sensor nodes attached to several hundred solar-powered sensor devices in the Springbrook National Park region, (bounded by 28 °14'1512" – 28 °13'138"S latitude, and 153°15'60" –153°16'43"E longitude) (See Figure 1). This tagged region is an area of approximately 0.13 square miles.

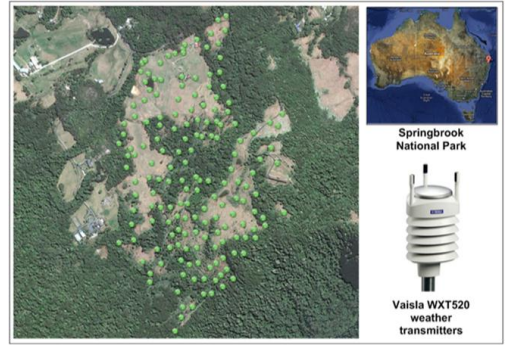

Figure 1. Study Area: the Springbrook WSN deployed in the Springbrook National Park

More specifically, in this project CSIRO scientists installed different types of sensor devices to monitor air temperature, relative humidity, wind speed, leaf wetness, soil water potential, total solar radiation, wind direction etc. The Springbrook sensor data set is ideal for our research because it records those parameters specifically required to calculate fire weather indices. For this study, we collected three types of weather variables: air temperature, wind speed and relative humidity from the Springbrook WSN project [6]. All the data were captured by Vaisla WXT520 weather transmitters. Each transmitter is sampled at 10 minute intervals.

### 3.2 Methodology

The proposed methodology for undertaking this research can be sub-divided into the following four major steps described below:

1. **Data Pre-processing and Storage**: A data pre-processing step was applied to detect and remove the outliers within the collected data streams. Then, the cleaned data streams were converted into RDF triples which are stored in optimized repository (multiple RDF triple stores).
2. **Semantic Inferencing**: 241 Fist-order logical rules were collected and converted to SPARQL inference rules by using the defined FWI ontology and other OWL ontologies. The SPARQL inference rules were saved in the optimized repository. Then, a rules-based inferencing algorithm was applied to generate fire weather indices and an Inverse Distance Weighting (IDW)-based neighbourhood region prediction algorithm was developed and applied to calculate more precise raster-based fire weather indices for a specific region at a time period (from point data).
3. **System Implementation**: A Web-based SFWI system was developed with a search interface to enable users to search fire weather indices for a specific region within a time period. A visualization interface consisting of Google Earth, timeline and pie charts was

   developed. The Google Earth map displays animated fire weather index search results, while the pie charts are used to compare entire period, day-time and night-time average fire weather indices.
4. **Evaluation**: Finally, three aspects of the system are evaluated. The first evaluation involves comparing our system with the McArthur Forest Fire Danger Index and the BoM Daily fire weather Index values. The second evaluation involves the assessment and optimization of SPARQL querying and RDF triple store configuration to obtain scalable performance. The third evaluation involves the assessment of the SFWI system usability by acquiring user feedback from 8 users.

# 4. DATA PRE-PROCESSING AND STORAGE
## 4.1 OUTLIER REMOVEMENT AND SEMANTIC ANNOTATION

Data quality is a significant issue associated with WSNs due to the limited resources available (power, memory, computational capacity and communication bandwidth), and the harsh environmental conditions [30-33]. In particular, outliers can significantly affect the accuracy of our fire weather index calculations. To detect outliers, we take advantage of the fact that air temperature, humidity and wind speed readings from sensors geographically close to each other, follow the same patterns [32,34]. In this study, we adopt an outlier detection approach, proposed by our previous work [35], to remove the outliers within the raw sensor data streams to generate the cleaned wireless sensor data streams.

Machine-processable semantics allows semi-autonomous or autonomous agents to streamline the capture, reasoning and re-use of data streams. To date, there exist a wide variety of existing sensor and observation ontologies to describe sensor networks, sensor devices and sensor observations with machine-interoperable semantics. Such ontologies include the Semantic Sensor Network Ontology (SSN) [36], OntoSensor [37], and CSIRO's Sensor Ontology [38]. The SSN ontology, developed by the W3C Semantic Sensor Network Incubator group, is the most comprehensive ontology which provides the top-level classes and properties for representing sensors, the measurement capabilities of sensors, and the sensor observations. More specifically, the SSN has defined the core concepts and relations (*sensors, features, properties, observations* and *systems*) and has been aligned to the DOLCE-Ultra Lite (DUL) ontology [39,40]. In addition, a number of groups have developed extensions to SSN, that we are able to take advantage of. The W3C Semantic Sensor Network Incubator group developed an Automatic Weather Station (AWS) ontology [41] to specify different sensor types, including *aws:TemperatureSensor, aws:WindSensor* and *aws:HumiditySensor.* The W3C Semantic Sensor Network Incubator group also developed a Climate and Forecast (CF) ontology [42] that defines climate data variables, such as *cf:air_temperature, cf:relative_humidity* and *cf:wind_speed.* NASA developed a unit ontology [43] that defines vocabularies for physical properties and corresponding units of measurements, such as *unit:degreeCelsius, unit:metrePerSecond* and *unit:percent* etc. Figure 2 illustrates how we combine these ontologies to describe our sensor data streams.

Before using SPARQL inferencing to estimate fire weather indices, we first need to convert the cleaned sensor observations (formatted in CSV files) to RDF triples. The combined SSN, AWS, DUL, and Unit ontologies were employed to convert the CSV formatted sensor observations to RDF triples. For instance, suppose that we have a temperature observation: "*2012-01-02 03:50:00, air_temperature, AT_1, SN_1, 13.5, °C"*. Parsing of this string is performed as follows: *2012-01-02 03:50:00* is the time when an observation was collected; *air_temperature* indicates the property of the observation; *AT_1* is the ID of the sensor device; *SN_1* is the ID of the sensor node that the sensor was attached to; *13.5* is the observation value; and the *°C* is the observation unit. After parsing, this observation is converted to RDF triples with spatial, temporal and semantic metadata (as shown in Figure 2).

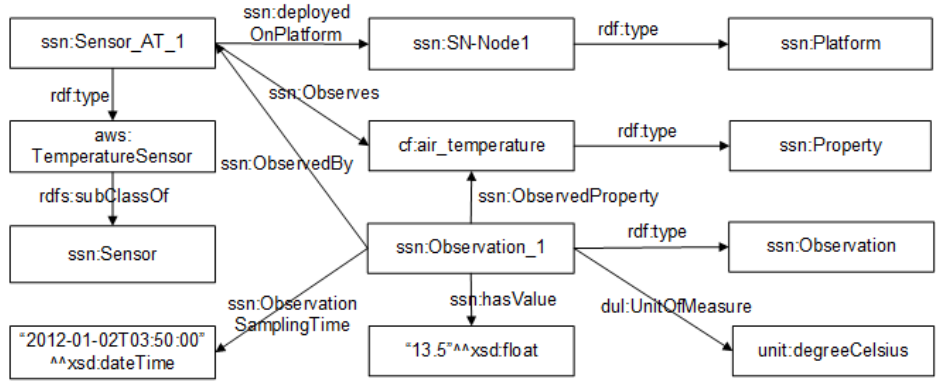

Figure 2. Instance of a temperature observation represented using the SSN, AWS, DUL, CF and Unit ontologies

```
Algorithm 1: Storing sensor observations to the multiple repository triple store
Input:  O indicates one type of sensor observation data set
        TP indicates the property of the sensor observation
Output: true or false. (If storage is successful, then return true. If storage failed, then return false)
        O1=preprocessing (O); //remove the outlier from the sensor observations
        RDFGraph =Conversion (O1); //convert the cleaned sensor observations to RDF graph
        Context=getContext(RDFGraph); //generate a context for the generated RDF graph
        RepositoryId=getRepositoryId( TP ); //look for the target repository Id
        R= SaveToRepository(RepositoryId , RdfGraph, Context);
                //save the RDF graphs to the corresponding repository
        If R==true, then
                maxTime=getMaxTime (RDFGraph); //get the maximum sampling time
                minTime=getMinTime (RDFGraph); // get the minimum sampling time
                UpdateCatalog (RepositoryId, Context, maxTime, minTime);
                //update the provenance information of the RDF graphs to the catalog repository
                Return true;
        else
                Return false;
        end
```

Figure 3. Algorithm for saving sensor observations to the multiple repository triple store

## 4.2 Storage Techniques

It is widely acknowledged that RDF graph-based triple stores are not very efficient in terms of query and reasoning performance [46], thus we propose a multiple repository storage approach to improve the query and reasoning performance of RDF graph-based triple stores. This multiple repository storage consists of a RDF graph catalog repository, and a set of RDF graph sub-repositories. The catalog repository is used to store catalog information about the sub-repositories (e.g., sub-repositoryID, sub-repository graphs, graph maximum Time, graph minimum Time, sub-repository data type etc.). There are four sub-repositories: relative humidity, air temperature, wind speed and a FWI repository. The relative humidity repository, wind speed repository and air temperature repository are used to store relative humidity RDF graphs, wind speed RDF graphs, and air temperature RDF graphs, respectively. When a user uploads a set of raw sensor observations to the system, the system will firstly filter the data using our data pre-processing approach and then convert them to RDF graphs. Next, the system automatically generates a unique ID for each graph (that defines the place/time/indicator context). Based on the uploaded data type, the system chooses a sub-repository to save this RDF graph. If the RDF graph is successfully uploaded to the repository, the system will save the corresponding graph storage information (context, maxTime and minTime, repositoryID) to the catalog repository to enable efficient searching and retrieval in the future. Figure 3 illustrates how to save raw sensor observations to a corresponding repository. The *FWI repository* is used to store the SPARQL inferencing results that calculate the fire weather indices. Detailed information describing the storage of inference results is explained in section 5.3.

# 5. SEMANTIC INFERENCING
## 5.1 Fire Weather Index Ontology

Next we developed a Fire Weather Index (FWI) ontology to define fire weather index classes and relative properties. Fire weather indices can be categorized into five high level categories: *fwi:Low, fwi:Moderate, fwi:High, fwi:VeryHigh* and *fwi:Extreme*. For the *fwi:Low* category, fires can be easily controlled and there will be no risk to life or forest. For the *fwi:Moderate* category, fires can be easily controlled but still present a threat. For the *fwi:High* category, fires can be controlled but present a threat. For the *fwi:VeryHigh* category, fires can be difficult to control and present a real threat, while for the *fwi:Extreme* category fires will likely be uncontrollable and fast moving with flames that may be higher than roof tops. Moreover, these five high level classes can be further sub-classed into 15 subclasses: (*fwi:Min-Low, fwi:Mid-Low, fwi:Max-Low), (fwi:Min-Moderate, fwi:Mid-Moderate, fwi:Max-Moderate), (fwi:Min-High, fwi:Mid-High, fwi:Max-High), (fwi:Min-VeryHigh, fwi:Mid-VeryHigh, fwi:Max-VeryHigh), (fwi:Min-Extreme, fwi:Mid-Extreme,* and *fwi:Max-Extreme)*. Finally, the FWI ontology was aligned to the Provenance (PROV) ontology [44], which provides a set of classes, properties, and restrictions to represent the provenance information. Figure 4 illustrates how the FWI ontology can be applied to describe a high fire weather event that was observed at Sensor Node 2.

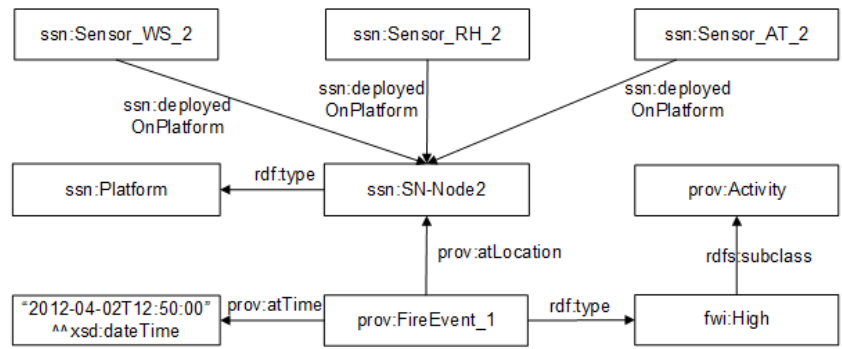

Figure 4. Applying the FWI ontology to describe the High fire weather index from three sensors (WindSpeed WS_2, RelativeHumidity RH_2, AirTemperature AT_2) on SN-Node2

## 5.2 Defining SPARQL Inference Rules

Inferencing is a mechanism by which a set of rules that represent domain expert knowledge are used to logically derive additional domain specific knowledge. SPARQL [14] is a standard RDF query language recommended by the World Wide Web Consortium (W3C) and considered as an essential graph-matching query language. More specifically, a SPARQL query consists of a body, a complex RDF graph pattern matching expression (which may include basic graph patterns, group graph patterns, optional graph patterns, alternative graph patterns and patterns on named graphs), and a head, an expression that indicates how to construct an answer to the query [45]. More specially, SPARQL has four types of query forms (Select, Construct, Ask and Describe) to define four types of result formats.

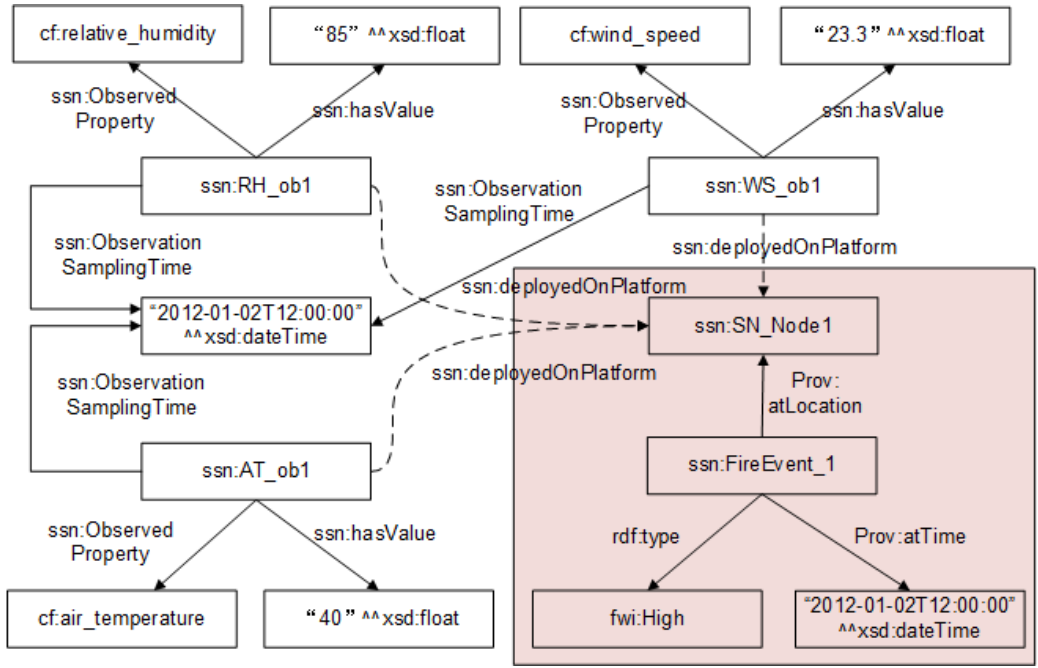

Figure 5. An example of applying SPARQL reasoning to an RDF dataset to infer a High fire weather index

SPARQL [14] can be used for inferencing because the SPARQL *Construct query form* returns an RDF graph that is formed by taking each query solution in sequence, substituting for the variables in the graph pattern, and combining the triples into a single RDF graph. In other words, the *Construct query form* can derive new triples when the graph patterns match. The returned RDF graph can be directly inserted into a repository by using the *Insert graph update* operation. Consider the following example using a SPARQL *Construct query form* and *Insert graph update* operation to infer a fire weather index value. In order to infer fire weather indices, we collected 241 fire weather index rules in total from the meteorologists. These rules were represented as SPARQL queries and then stored in the multiple repository storage. For example, one of these rules is expressed as: "if relative humidity >= 80% AND 17.5 m/s <= wind speed <=24.4m/s AND 32 °C <= air temperature <= 41 °C, for a given location at time T, then at time T this location has a *High* fire weather index." We can interpret this rule statement into a SPARQL rule expressed as follow:

```
Construct {
    ?FireEvent_1 prov:atLocation ?node.
    ?FireEvent_1 prov:atTime ?T.
    ?FireEvent_1 rdf:type fwi:High.
}

Where{
    ?RH_OB1 ssn:ObservedProperty cf:relative_humidity.    ?RH_OB1 ssn:ObservationSamplingTime ?T.
    ?RH_OB1 dul:unitOfMeasure unit:percent.               ?RH_OB1 ssn:ObservedBy ?RH_Sensor1.
    ?RH_Sensor1 ssn:deployedOnPlatform ?node.             ?RH_Sensor1 ssn:hasValue ?RH_OB1V.
    ?WS_OB1 ssn:ObservedProperty cf:wind_speed.           ?WS_OB1 ssn:ObservationSamplingTime ?T.
    ?WS_OB1 dul: unitOfMeasure unit:meterPerSecond.       ?WS_OB1 ssn:ObservedBy ?WS_Sensor1.
    ?WS_Sensor1 ssn:deployedOnPlatform ?node.             ?WS OB1 ssn:hasValue ?WS_OB1V.
    ?AT_OB1 ssn:ObservedProperty cf:air_temperature.      ?AT_OB1 ssn:ObservationSamplingTime ?T
    ?AT_OB1 dul: unitOfMeasure unit:degreeCelsius         ?AT_OB1 ssn:ObservedBy ?AT_Sensor1.
    ?AT_Sensor1 ssn:deployedOnPlatform ?node.             ?AT_Sensor1 ssn:hasValue ?AT_OB1V.
    Filter(
    ?RH_OB1V>=80&&?RH_OB1V<=100&&
    ?WS_OB1V>=17.5&&?WS_OB1V<=24.4&&
    ?AT_OB1V>=32&&?AT_OB1V<=41)
}
```

Suppose that we collected three type of sensor observations (relative humidity: 85%, wind speed: 23.3m/s, and air temperature: 40°C) from the sensor node 1 at 2012-01-02T12:00:00, then these sensor observations can be expressed in RDF format as shown in Figure 5. When we apply the above inference rule to this data set, then it generates new triples (shown in Figure 5 shaded pink), which are the inference results that are saved to the *FWI Repository*.

```
Algorithm 2: Inferring fire weather index
Input: [t1,t2]
Output: Fire Weather Indices (FWIs) within the period [t1,t2]
hasAll=CheckFW(FWIRepository, t1, t2);
if hasAll==false, then
        timeRanges=getMissedTimeRanges(FWIRepository, t1, t2);
        temporaryRepository=createATemporaryRepository();
        repositoryIds={airTemperatureRepoId, windSpeedRepoId, relativeHumidityRepoId};
        for each repositoryId in the repositoryIds list, do
                graphIds=getRDFGraphsStorageInfor(CatalogRepository, repositoryId, timeRanges);
                RDFTriples=getRDFTriplesFromRepo(RepositoryId, graphIds, timeRanges);
                SaveRDFTriplesToTemporaryRepository(RDFTriples);
        end for
        newFWIs=inferringFWIsAtTemporaryRepository(SPARQLRules, temporaryRepository);
        SaveFWIs(FWIRepository, newFWIs);
        destoryTemporaryRepository(temporaryRepository);
end if
FWIs=searchFWIs(FWIRepository, t1, t2);
Return FWIs;
```

Figure 6. Inference algorithm for inferring new FWIs

### 5.3 Inference Algorithm

The Canadian Forest Fire Weather Index system takes Noon Stand Time weather parameters as input. Hence, it only produces the daily fire weather indices at noon local time, which limits its accuracy and usefulness in decision-making. Our approach (Figure 6) enables users to access and query the fire weather indices in real time and at 10 minute intervals. Moreover, our system only performs the inferencing operation when it receives a query from a user. When users input their query time parameters to the system, the system will search the *FWI Repository* which stores all the historical inferred fire weather indices. If the *FWI Repository* contains the requested fire weather indices then it will return them. If not, the system will retrieve the time range information for the missing fire weather indices and create a temporary RDF graph repository as a data hub to estimate fire weather indices. Next, the system visits all the sub-repositories to upload related sensor data combinations to the temporary repository. Then, a SPARQL inference operation applies the inferencing rules to infer all the fire weather indices. Finally, the inferred fire weather indices are saved in the *FWI Repository* and the requested fire weather index results are presented to the user.

### 5.4 IDW-Based Neighborhood Region Prediction

In Section 4.1 we described the pre-processing step to remove outliers within the sensor data streams. However, data loss often occurs in WSNs due to random link faults, hazard node faults, inaccuracies of measurements, calibration errors, and fading signal strength etc. [9,10]. Thus, an additional goal of this study is to reduce inaccuracies of inference results that occur due to data loss. Therefore, we propose a neighborhood region prediction approach based on assumptions that the fire weather index of a specific place is influenced mostly by the nearby sensor nodes and less by the more distant sensor nodes. This approach utilizes a commonly used technique – Inverse Distance Weighting (IDW) [15] for interpolation between a known scattered set of points. Let $f(sn_i)$ be the fire weather index at sensor node $sn_i$, $\{f(sn_1), f(sn_2), \cdots, f(sn_N)\}$ be the fire weather indices set, and $N$ be the total number of fire weather indices. Hence, we can use the following formula to calculate the fire weather index $f(L)$ at a specific location $L$:

$$f(L) = \sum_{i=1}^{N} \frac{w_i(L) f(sn_i)}{\sum_{j=1}^{N} w_j(L)} \qquad (1)$$

where $i$ ($1 \leq i \leq N$), $j$ ($1 \leq j \leq N$), and

$$w_i(L) = \frac{1}{d(sn_i, L)^p} \qquad (2)$$

Where $p$ is the power parameter (typically, $p = 2$), and $d(sn_i, L)$ is the distance between sensor node $sn_i$ and location $L$.

$$d(sn_i, L) = r * a\cos\left(\sin(x) * \sin(x_i) + \cos(x) * \cos(x_i) * \cos(y - y_i)\right) \qquad (3)$$

Where $x$ and $y$ are the latitude value and longitude value of the location $L$, $x_i$ and $y_i$ are the latitude value and longitude value of the sensor node $sn_i$, and $r$ (typically, $r = 6373.8$ kilometers) is the radius of the Earth. This IDW-Based Neighbourhood Region Prediction algorithm is applied to the inferred fire weather indices, prior to the display and animation of sFWI spatial distributions via the Google Earth mapping interface (described in Section 7.2).

## 6. IMPLEMENTATION
### 6.1 System Architecture

Figure 7 depicts the principle technical components of our Semantic Fire Weather Index (SFWI) system and the data flow between them. Our architecture combines programming technologies (Java, JavaScript, and JSON) and Web-based visualization technologies (Google Earth, Keyhole Markup Language (KML), timeline and pie charts) to maximize access and interactivity. In addition, we apply Semantic Web technologies (RDF, OWL ontologies, SPARQL, and RDF Sesame Repository Stores) to enrich sensor data with domain-specific semantic metadata, to reason across the sensor data streams, and to discover implicit knowledge such as fire weather index. More specifically, we adopt Java IDE Eclipse and Semantic Web plugin Sesame to support the development.
A large volume of micro-climate sensor data collected by Vaisala Weather Transmitter WXT520 devices is harvested from the Springbrook WSN project: including 2.5 years of relative humidity sensor data, air temperature sensor data and wind speed sensor data. The data processing step performs data cleaning (implemented by Java) to remove the outliers and conversion to RDF (implemented in Java). The generated RDF triples are stored in the relative humidity, air temperature and wind speed repositories respectively. The inference component performs the SPARQL reasoning to infer the fire weather indices at specific points/sensor nodes and saves the results in the FWI repository.

A search interface was developed to enable users to search and retrieve fire weather index from the FWI Repository by specifying a region and time period. After a query region is specified, the IDW-Based Neighbourhood Region Prediction is applied to generate raster-based visualizations dynamically displayed in a Web-based Google Earth interface. In addition, a browser-based timeline to display the fire weather trends over the query period (at 10 min intervals) and three pie charts to compare average fire weather indices for the entire period, day time and night time.

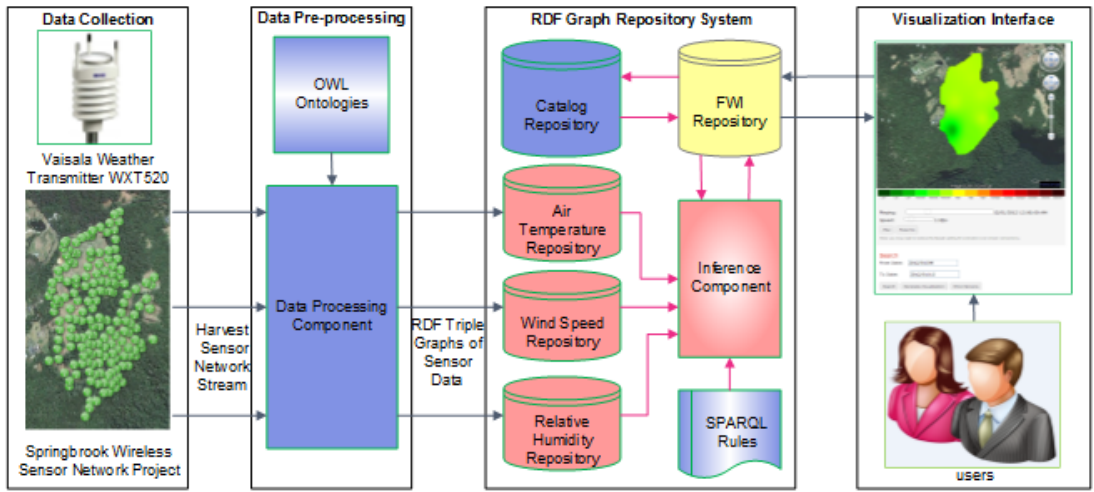

Figure 7. High level architecture view of the SFWI system

## 6.2 User Interface

The user interface (shown in Figure 8) (accessible via a Firefox or Chrome Web browser), enables users to interactively:

- Search and retrieve the fire weather indices for a specific region and time range. For example, show me the fire weather indices between "09/01/2012 12:00:00" and "13/02/2012 12:00:00". The search results are displayed overlaid on the specified region using Google Earth layers (See Figure 8 (a)).
- Statistically analyze the fire weather indices to compare day-time and night-time results. The results are displayed as pie charts (See Figure 8 (b)).

The Google Earth animations dynamically show how FWI values vary over time. The pie charts on the right show the statistical information about the average, day-time and night-time fire weather indices. From the pie chart, we can affirm that typically FWIs are higher during daylight hours (15% low-, 62.5% low, and 22.5% moderate) than during night time (64.7% low-, 15.7% low, 2% low+, 5.9% moderate+ and 11.8% high-). However, the pie chart also illustrates that although night time is more likely to show lower fire weather indices, in some cases, higher fire weather indices may appear during the night.

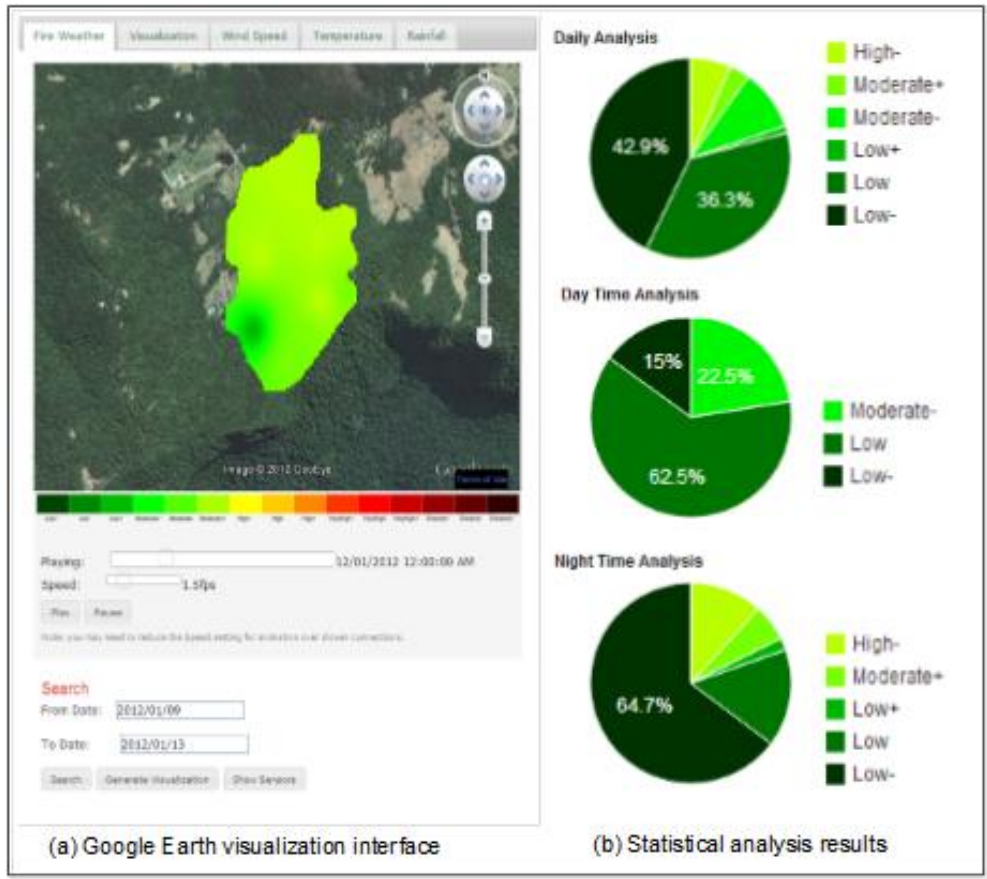

Figure 8. User interface for Displaying SFWI Search, Browse results

# 7. EVALUATION
## 7.1 Evaluation of the Accuracy and Precision of Fire Weather Index Calculations

We evaluate the accuracy and precision of our approach, by comparing our results with the McArthur Forest Fire Danger Index (FFDI) and the Bureau of Meteorology's (BoM) Daily Fire Weather Index (which is based on the Canadian Fire Weather Index).

Firstly, we collected 6 months of data (from 01/01/2012 00:00:00 to 01/07/2012 00:00:00) from the Springbrook project. These data sets consist of three weather parameters: air temperature, relative humidity and wind speed. This assessment involved calculating the similarity between our system fire weather indices estimates and the FFDI estimates for the same region. The similarity results for each month are presented in Figure 9. They reveal that our method and the FFDI produced similar results: Jan.2012 similarity $\geq$98%, Feb. 2012 similarity $\geq$99%, Mar. 2012 similarity $\geq$96%, Apr. 2012 similarity $\geq$97%, May similarity $\geq$99.5%, and Jun. 2012 similarity $\geq$99%. These results prove that our inference rules (as defined by the domain experts) can accurately estimate fire weather indices.

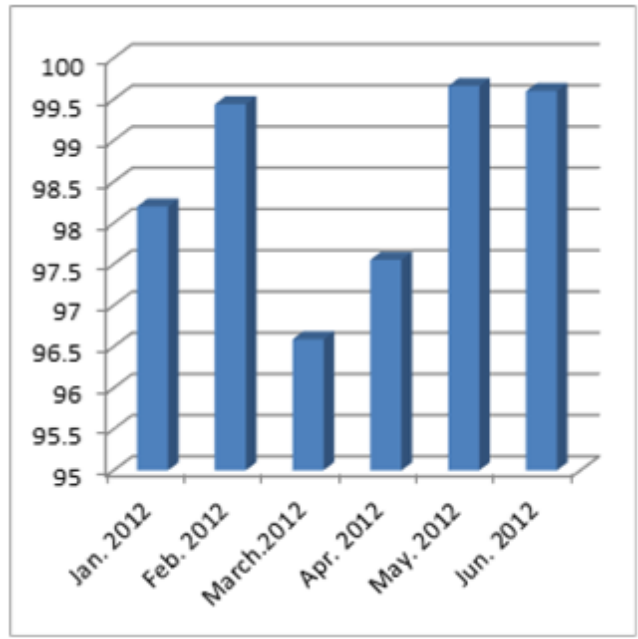

Figure 9. Similarities between FFDI estimates and SFWI estimates for period Jan- Jun 2012

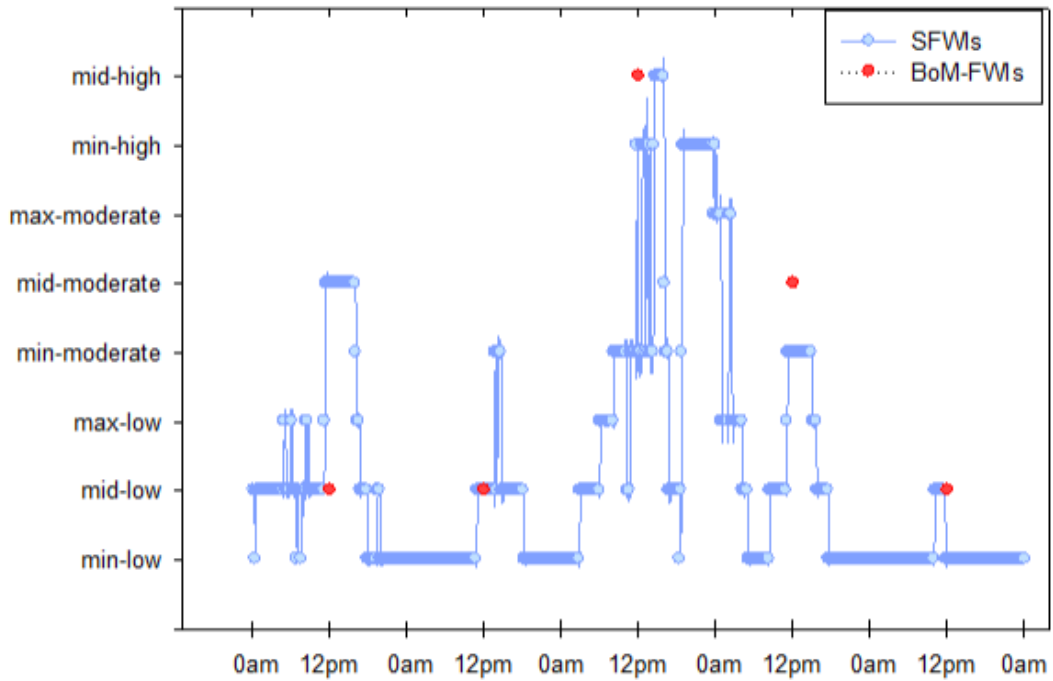

Figure 10. Comparison of BoM-FWIs with SFWIs for period 09/01/2012 00:00:00 - 14/01/2012 00:00:00

Secondly, we compared the BoM's Daily Fire Weather Indices (BoM-FWIs) with our semantic-based high resolution dynamic Fire Weather Indices (SFWIs) (that changes every 10 minutes) for a 5 day period (from 09/01/2012 00:00:00 to 14/01/2012 00:00:00). The aim of this assessment is International Journal of Web & Semantic Technology (IJWesT) Vol.5, No.4, October 2014to evaluate the precision of our calculations. The evaluation results, shown in Figure 10, reveal that our approach is able to provide more precise results. The results also show that FWI values have a strong relationships with time. More specifically, FWI values in the early afternoon are higher than in any other period. In addition, Figure 10 demonstrates that our system is able to infer more accurate and precise fire weather index values with finer spatio-temporal resolution than the traditional approaches.

## 7.2 Evaluation of the RDF Triple Stores and SPARQL Query Performance

We also evaluated the performance of the SPARQL inference and query engine by measuring the speed of SPARQL query and inference. 144,966 triples of weather parameters were generated. Next, we stored a copy of the triples in one RDF repository (1R) which is the typical approach and a copy of the triples in our multiple repository storage (MR) approach. To conduct this evaluation, we executed 8 new SPARQL queries over different time periods (e.g. 1 hour, 6 hours etc.) on the single RDF (1R) repository to infer fire weather indices. The corresponding execution time results are calculated and presented in Figure 11 labelled (NQ-1R) (New Query – 1 Repository). Secondly, we re-ran these 8 SPARQL queries again on the single RDF repository. The execution time results presented in Figure 11 are labelled as (RQ-1R) (Repeat Query – 1 Repository). Thirdly, we ran these 8 new SPARQL queries on the multiple repository storage to infer the corresponding fire weather indices. The execution time results are presented in Figure 11 and labelled as (NQ-MR) (New Query – Multiple Repository). Lastly, we re-ran these 8 SPARQL queries again on the multiple repository storage. The execution time results are presented in Figure 11 labelled as (RQ-MR) (Repeat Query- Multiple Repository). The overall results illustrated in Figure 11 show that our multiple repository storage outperforms the single repository approach. Moreover, our method performs significantly better than the single repository approach in terms of running repeated queries over a long period. Therefore, our approach demonstrates greatly improved querying speed.

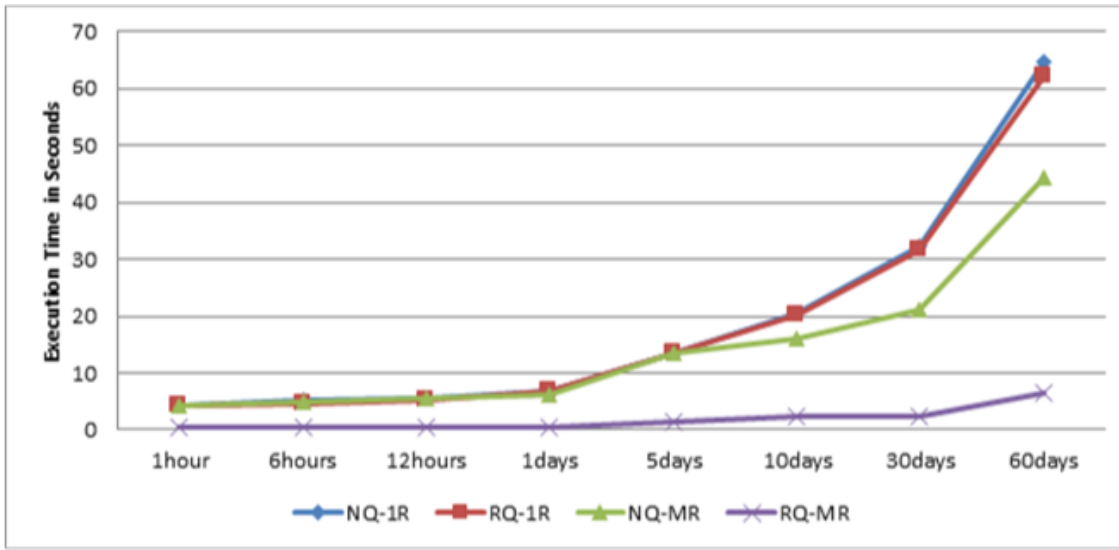

Figure 11. Execution time for running SPARQL queries over different time periods for new queries and repeat queries over single and multiple repositories respectively (NQ-1R, RQ-1R, NQ-MR, RQ-MR)

International Journal of Web & Semantic Technology (IJWesT) Vol.5, No.4, October 2014### 7.3 Evaluation of System Usability

Finally, the system usability was assessed by soliciting feedback from eight users via a questionnaire and by observing user behaviour during the tests. Eight users were asked to respond to the following questions on the questionnaire after they had completed a given set of tasks using the system:

- Q1 - I found the user interface for searching/querying fire weather indices easy to use;
- Q2 - I found the Google Earth visualization and animations useful;
- Q3 - I found the timeline visualization useful;
- Q4 - I found the daily, day time and night time charts useful;
- Q5 - I could understand the spatial and temporal patterns of fire weather indices better after using the system.

Users were asked to provide a response to each question from a five-point Likert scale: 1=Strongly agree; 2=Agree; 3=Neither agree nor disagree; 4=Disagree; 5=Strongly disagree.

| Questions | Q1 | Q2 | Q3 | Q4 | Q5 |
|---|---|---|---|---|---|
| Positive Feedback | 100% | 100% | 87.5% | 100% | 100% |

Table 1. The evaluation results of usability

User feedback from the questionnaire was presented in Table 1 and on-the-whole positive. All the users found that the user interface was intuitive and easy to use, and the Google Earth visualizations and the pie charts visualizations useful for exploring variations and trends in fire weather indices over time and location. Users requested that the timeline visualization should be improved to support zoom in and zoom out functionalities. They felt this would be more useful for meteorologists or regional fire safety agencies to understand the fire weather index fluctuation trends over a long period or between regions. Moreover, a significant number of users requested the ability to overlay the Google Earth visualization fire weather index map with country roads, Digital Elevation Models (DEM) and land cover maps (forest cover or grassland). They felt that this would greatly assist firefighters and/or residents in developing fire fighting, evacuation and rescue plans.

## 8. FUTURE WORK AND CONCLUSIONS

In conclusion, many micro-climate WSNs collect observations and measurements about environment physical parameters which provide ideal data for estimating bushfire risks. However, the quality of WSN data largely depends on the configuration of the sensor network. Existing fire weather index analyses that take noon time weather parameters collected from the distributed weather stations as input, cannot provide residents, firefighters or fire officials with accurate fire weather index information for a local region. Hence, we developed the Semantic Fire Weather Index (SFWI) system which overcomes the limitations of the existing fire weather index by estimating fire weather index from micro-scale WSN data streams. Specifically, we integrated semantic reasoning with domain expert knowledge to estimate fire weather index for a specific region and time. Data pre-processing and an IDW-based algorithm were developed to support and improve the accuracy of the data underpinning our system. A FWI ontology was developed to enable the description of different level fire weather indices. In addition, we developed a visualization interface that enables fire managers, and the general public to access the fire weather index data via an easy-to-use mapping and timeline interface. The outcome is an extensible framework and a robust foundation for future advanced WSNs that can be used to enhance the development of fire rescue system or other environmental decision support systems.

Future research is required to integrate SPARQL inference with fuzzy inference technology to provide users with estimates of uncertainty together with the fire weather indices. In addition, a user interface is required to enable meteorologists to easily enter, save and publish their domain expert knowledge via human-readable rules that can be translated to corresponding SPARQL rules. Finally, other environmental data sets such as land cover (forest land cover and grassland land cover), digital elevation models, creek and road networks could usefully be combined with the sensor network data streams to assist firefighters, local residents and natural disaster managers in decision-making.

## ACKNOWLEDGMENTS

The work presented in this paper is supported by the China Scholarship Council. The authors would like to thank Mr Luke Hovington and Mr Darren Moore from the CSIRO ICT Center for their valuable support and contributions and feedback.

International Journal of Web & Semantic Technology (IJWesT) Vol.5, No.4, October 2014